\documentclass[11pt]{article} 
 
\begin{document} 
 
\title{Formation and displacement of bubbles in a packed bed  \\ Fluid Dynamics Videos} 
 
\author{ Enrique Soto, Alicia Aguilar-Corona* and Roberto Zenit\\ 
\\\vspace{6pt} Universidad Nacional Autonoma de Mexico, Mexico \\ *Universidad Michoacana de San Nicolas de Hidalgo, Morelia, Mich, Mexico} 
 
\maketitle 
 
 
\begin{abstract}


The fluid dynamics video show a gas stream which is injected into a packed bed immersed in water and fluid dynamcis video present the dynamics 

involved (\href{http://ecommons.library.cornell.edu/bitstream/1813/8237/2/LIFTED_H2_EMS
T_FUEL.mpg}{Video 
1} and 
\href{http://ecommons.library.cornell.edu/bitstream/1813/8237/4/LIFTED_H2_IEM
_FUEL.mpg}{Video 
2}). The refractive index of the water an the packed bed are quite similar and the edges of the spherical 

particles can be seen. Two distinctive regimens can be observed. The first one, for low air flow rates, which is 

characterized by the percolation of the air thought the interstitial space among particles. And the second one, 

for high air flow rates, which is characterized by the accumulation of air inside the packed bed without 

percolation, it can be observed that the bubble pull apart the particles apart. Furthermore, for the first case 

the position of the particles remains constant while for the second one a circulation of particles is induced by 

the bubbles flow.

\end{abstract} 
\section{References} 
\begin{enumerate} 
\item Gostiaux, L., Gayvallet,H. and Géminard, J.-C. 2002 Dynamics of a gas bubble rising
through a thin immersed layer of granular material: an experimental study. GranularMatter, 4,39-44.
 
\end{enumerate} 
\end{document}